\documentclass[12pt,preprint]{aastex}
\newcommand{\beq}{\begin{equation}}
\newcommand{\beqa}{\begin{eqnarray}}
\newcommand{\eeq}{\end{equation}}
\newcommand{\eeqa}{\end{eqnarray}}
\newcommand{\simg}{\gtrsim}
\newcommand{\siml}{\lesssim}
\newcommand{\meszaros}{M${\acute {\rm e}}$sz${\acute {\rm a}}$ros}
\shorttitle{Variabilities of GRB afterglows}
\shortauthors{Ioka, Kobayashi, \& Zhang}
\begin{document}
\title{
Variabilities of Gamma-Ray Burst Afterglows:
Long-acting Engine, Anisotropic Jet or Many Fluctuating Regions?
}
\author{
Kunihito Ioka\footnote{
Physics Department and Center for Gravitational Wave Physics, 
104 Davey Laboratory, Pennsylvania State University, University Park,
PA 16802}, 
Shiho Kobayashi$^{1,}$\footnote{
Department of Astronomy and Astrophysics, 525 Davey Laboratory,
Pennsylvania State University, University Park, PA 16802},
and Bing Zhang\footnote{
Department of Physics, University of Nevada, Las Vegas, NV 89154}
}

\begin{abstract}
We show that simple kinematic arguments can give limits on the timescale 
and amplitude of variabilities in gamma-ray burst (GRB) afterglows,
especially when the variability timescale is shorter than
the observed time since the burst $\Delta t < t$.
These limits help us to identify the sources of afterglow 
variability. The afterglows of GRB 011211 and GRB 021004 marginally 
violate these limits. If such violation is confirmed by the 
Swift satellite, a possible explanation is that (1) the compact 
objects that power GRB jets continue to eject an intermittent 
outflow for a very long timescale ($\simg 1$day), (2) the GRB 
jet from the central engine has a temporal anisotropy with a large 
brightness contrast $\simg 10$ and small angular structure $\siml 10^{-2}$,
or (3) many ($\simg 10^{3}$) regions fluctuate simultaneously 
in the emitting site.
\end{abstract}


\section{Introduction and summary}

The standard synchrotron shock model has been successful in
explaining overall features of the gamma-ray burst (GRB) afterglows
\citep[e.g.,][]{zhang04,meszaros02}.
The standard afterglow model assumes a single relativistic blast wave
expanding into an ambient medium with a spherical, smooth density
distribution.  The emitting surface (shock front) is assumed to be
homogeneous and spherical.  Such a model predicts smooth afterglow
light curves.

However, recent dense monitoring of afterglow lightcurves indicates that
at least some afterglows deviate from 
a smooth power law decay, such as in 
GRB 021004 \citep[e.g.,][]{lazzati02,kobayashi03,fox03,heyl03,nakar03}
and GRB 030329 \citep[e.g.,][]{uemura03,lipkin04,torii03,sato03,urata04}.
Also short-term variabilities (with the variability timescale $\Delta t$
shorter than the observed time since the burst $t$)
are observed in the afterglows of GRB 011211 \citep{jakobsson04,holland02}
and GRB 021004 \citep{bersier03,halpern02}.
These variabilities carry a wealth of information about
the central engine and its surroundings.

So far four major scenarios have been 
proposed for afterglow variabilities.
(1) The ambient density into which a blast wave
expands may have fluctuations (see \S~\ref{sec:den}).
(2) The emitting surface may have an intrinsic angular structure, i.e.
the so called patchy shell model (see \S~\ref{sec:patchy}).
(3) The shocks may be ``refreshed'' by slow shells that catch up with
the decelerated leading shell
(see \S~\ref{sec:refresh}). 
(4) The central engine may be still active at the observing time
(see \S~\ref{sec:long}).

In this paper, we show that some kinds of afterglow variability
are kinematically forbidden
under some standard assumptions.
We assume that the standard model determines the power-law baseline
of the afterglow flux $F_{\nu}$, 
and derive the following limits for dips (bumps)
that deviate below (above) the baseline with a timescale $\Delta t$
and amplitude $\Delta F_{\nu}$:
\begin{enumerate}

\item[(a)] No dips in afterglow light curves can have
a larger amplitude than the limit given by 
equation (\ref{eq:V-2}) (see \S~\ref{sec:negative}).

\item[(b)] Ambient density fluctuations cannot make a bump
in afterglow light curves larger than that in equation (\ref{eq:den2})
(see \S~\ref{sec:den}).

\item[(c)] Patchy shells cannot make a bump with a timescale shorter
than the observed time $\Delta t < t$,
although the rising time $\Delta t_{\rm rise} < t$ is allowed
(see \S~\ref{sec:patchy}).

\item[(d)] Refreshed shocks cannot make a bump with $\Delta t < t/4$
(see \S~\ref{sec:refresh}).

\end{enumerate}
These limits are summarized in Figure~\ref{fig:f1}.  Notice that the
limits (c) and (d) are derived from purely geometrical arguments, and
hence, they only give constraints on $\Delta t$ (not on $\Delta
F_{\nu}$).
When many regions fluctuate simultaneously,
the limits (a) and (b) are modified and are given by
equations (\ref{eq:manydip}) and (\ref{eq:manyden}), respectively, as discussed
in Appendix \S~\ref{sec:many}.

The observed variabilities in the afterglows of GRB 011211
\citep{jakobsson04,holland02} 
and GRB 021004 \citep{bersier03,halpern02}
may actually violate all the above limits (a)-(d),
although these are still within errors. We suggest\footnote{
\citet{jakobsson04} have also suggested the long-acting engine and 
the asymmetric density or energy variations as the origin of
the variabilities observed in GRB 011211 through geometrical considerations.} 
that this may imply that 
\begin{enumerate}
\item[(1)] the central engine is still active at the observed
time ($\simg 1$ day; see \S~\ref{sec:long}),

\item[(2)] the GRB jets have a temporal anisotropy
with a small angular structure and large brightness contrast
[equations (\ref{eq:dth}) and (\ref{eq:di}) in
\S~\ref{sec:dis}],

\item[(3)] or many ($\simg 10^{3}$) regions
fluctuate simultaneously in the emitting site
(see \S~\ref{sec:many}).
\end{enumerate}
Therefore variabilities in afterglow light curves may provide
important clues to the nature of the compact object that triggers the burst
and its surroundings.
The Swift satellite will significantly increase such samples,
and it will allow us to further understand the nature of GRB engines.

\section{Dips in afterglow light curves}\label{sec:negative}

First let us consider dips in afterglow light curves
at which the flux temporarily decreases below
the expected power-law decay 
with a duration $\Delta t$ smaller than the observing time $t$, i.e.,
$\Delta t \le t$.
We assume that dips arise from nonuniformity on the emitting surface
induced by a certain mechanism (e.g., density fluctuations).
Because of the relativistic beaming, the visible half-angular size
of an emitting surface with a Lorentz factor $\gamma$ is about $\gamma^{-1}$.
Since the emitting surface with a radius $\sim R$ has a curvature,
two photons emitted on the line-of-sight and at the edge of the visible surface
travel in different time to the observer.
This angular spreading time mainly determines the observed time since
the burst, i.e., $t \sim R/2 \gamma^{2} c$ \citep{fenimore96,sari97}.
If a variable region has a half-angular size $\Delta \theta (\le
\gamma^{-1})$ and 
a viewing angle (with respect to the center of the variable region)
$\theta_{v} (\le \gamma^{-1})$ to the observer, 
the angular spreading effect also puts a lower limit on
the variable timescale\footnote{
Note that $\Delta t \ge R[(\theta_{v}+\Delta \theta)^{2}-
(\theta_{v}-\Delta \theta)^{2}]/2 c= 2 R \Delta \theta \theta_{v}/c$
for the off-axis case.
}
as
$\Delta t \ge R \Delta \theta \max[\Delta \theta/2, 2 \theta_{v}]/c$.
(This does not depend on the Lorentz factor of the variable region.)
Then the ratio of the variable area 
$\Delta S \sim \pi (R \Delta \theta)^{2}$ to
the visible area $S \sim \pi (R/\gamma)^{2}$ has an upper limit,
\beqa
\frac{\Delta S}{S} \sim (\gamma \Delta \theta)^{2}
\le \left\{
\begin{array}{ll}
\Delta t/t,& ({\rm on-axis})\\
(1/4)(\Delta t/t)^{2},& ({\rm off-axis})
\end{array}\right.
\label{eq:S-}
\eeqa
where we use the typical value $\theta_{v}\sim \gamma^{-1}/2$ 
for the off-axis case $\theta_{v} \simg \Delta \theta$.

In addition, the thickness of the blast wave affected by a certain mechanism
(e.g., density fluctuations) should be 
less than the variability timescale multiplied by the speed of light 
$\sim c \Delta t$. Since the overall thickness of the blast wave is about
$\sim R/16 \gamma^{2} \sim c t/8$, the ratio of the variable volume 
$\Delta V \sim c \Delta t \Delta S$ to the visible volume $V \sim c t S/8$ 
is less than
\beqa
\frac{\Delta V}{V}
\le \left\{
\begin{array}{ll}
8 (\Delta t/t)^{2},& ({\rm on-axis})\\
2 (\Delta t/t)^{3}.& ({\rm off-axis})
\end{array}\right.
\label{eq:volume}
\eeqa
To obtain the upper limits on the amplitude of a dip,
we assume that the emission from the variable volume is suddenly shut off.
Additional effects (e.g., cooling timescale) only make the dip
less significant.
Then the kinematical upper limits for the dips are
\beqa
\frac{|\Delta F_{\nu}|}{F_{\nu}}
\le \frac{\Delta V}{V}
\le \left\{
\begin{array}{ll}
8 (\Delta t/t)^{2},& ({\rm on-axis})\\
2 (\Delta t/t)^{3},& ({\rm off-axis})
\end{array}\right.
\label{eq:V-}
\eeqa
regardless of the cause of the variability,
as long as the dips are produced by a disturbance on
the emitting surface.

When deriving equation (\ref{eq:V-}) from equation (\ref{eq:volume}),
we assumed that the visible region has a uniform brightness.
In reality a spherical afterglow appears on the sky as a ring
because of the relativistic effect
\citep{pana98,sari98,waxman97}.
Since the surface brightness normalized by its mean value
is about $I_{\nu}/\bar I_{\nu} \sim 0.1$ at the center and
about $I_{\nu}/\bar I_{\nu} \sim 3$ 
at the edge in the optical band $(F_{\nu} \propto \nu^{(1-p)/2})$
even for a uniform surface \citep{granot01},
we replace equation (\ref{eq:V-}) by
\beqa
\frac{|\Delta F_{\nu}|}{F_{\nu}}
\le \left\{
\begin{array}{ll}
(4/5)(\Delta t/t)^{2},& ({\rm on-axis})\\
6 (\Delta t/t)^{3}.& ({\rm off-axis})
\end{array}\right.
\label{eq:V-2}
\eeqa

The above limit is applicable when the emitting surface has one single 
variable region.
When many regions fluctuate simultaneously, 
one may use equation (\ref{eq:manydip}) instead, 
as discussed in \S~\ref{sec:many}.
Notice, however, that the limit in equation (\ref{eq:V-2}) is still
useful since its violation implies that there are many variable
regions.

\section{Bumps in afterglow light curves}

Since the surface brightness of variable regions
has the lower limit (zero),
we could give constraints on dips in a general way.
However, when considering afterglow bumps, since the surface
brightness of variable regions has no upper limit in principle,
its constraint should depend on the specific model of bumps.
In the following we separately discuss each probable scenario
in turn.

\subsection{Ambient density fluctuations}\label{sec:den}

Ambient density fluctuations may lead to afterglow variabilities
\citep{wang00,lazzati02,dai02,nakar03}.
Such fluctuations may be due to turbulence of interstellar medium 
or variable winds from the progenitor star.
Here we obtain a kinematical upper limit on the variabilities
that does not depend on the properties of the density fluctuations.

As in the case of dips, which we have discussed in the previous
section, the ratio of the variable volume $\Delta V$ 
to the visible volume $V$ satisfies equation (\ref{eq:volume}).
Since the kinetic energy
in the visible volume $E_{\rm kin}$ is almost uniformly distributed
and the conversion efficiency from the kinetic energy $E_{\rm kin}$ 
to the internal energy $E$ (lab-frame) by shocks is close to unity 
(i.e., $E \sim E_{\rm kin}$) if the shocks are relativistic,
we have $\Delta E/E \siml \Delta E_{\rm kin}/E_{\rm kin} \sim \Delta V/V$ 
where $\Delta E (\siml \Delta E_{\rm kin})$
is the internal (lab-frame) energy to produce the variability.
Therefore the bolometric luminosity ratio of the variable part $\Delta L$
to the base level $L$ is less than
\beqa
\frac{\Delta L}{L} \le \frac{\Delta E/\Delta t}{f_{c} E/t}
\le \left\{
\begin{array}{ll}
8 f_{c}^{-1} (\Delta t/t),& ({\rm on-axis})\\
2 f_{c}^{-1} (\Delta t/t)^{2},& ({\rm off-axis})
\end{array}\right.
\label{eq:L}
\eeqa
where we assume $L \sim \epsilon_{e} f_{c} E/t$, which is a good approximation
for the standard afterglow, and $\Delta L \le \epsilon_{e} \Delta E/\Delta t$,
which does not depend on the precise radiative process.\footnote{
We implicitly assume that the energy fraction that goes into
electrons $\epsilon_{e}$ is determined by the microscopic physics
and it is constant.}
The fraction of cooling energy is about
$f_{c} \sim (\nu_{m}/\nu_{c})^{(p-2)/2} \sim 1/2$ 
for the typical standard afterglow, i.e.,
the cooling frequency $\nu_{c} \sim 10^{15}$ Hz,
the characteristic synchrotron frequency $\nu_{m} \sim 10^{12}$ Hz,
and the electron power-law distribution index $p \sim 2.2$
\citep{sariea98}.

Since the variable flux $\nu \Delta F_{\nu}$ at an observed frequency $\nu$
is clearly less than the bolometric variable flux 
$\Delta F$ (i.e., $\nu \Delta F_{\nu} \le \Delta F$),
we can put the upper limits on bumps
due to density fluctuations as
\beqa
\frac{\Delta F_{\nu}}{F_{\nu}}
\le \frac{F}{\nu F_{\nu}}
\frac{\Delta F}{F}
\le \left\{
\begin{array}{ll}
8 f_{c}^{-1} (F/\nu F_{\nu}) (\Delta t/t),& ({\rm on-axis})\\
2 f_{c}^{-1} (F/\nu F_{\nu}) (\Delta t/t)^{2},& ({\rm off-axis})
\end{array}\right.
\label{eq:den}
\eeqa
where $F$ is the bolometric base flux.
The second inequality in equation (\ref{eq:den}) 
was derived by using $\Delta F/F \le \Delta L/L$ and equation (\ref{eq:L}).
This is because the bolometric flux $F$ ($\Delta F$) 
is proportional to the luminosity $L$ ($\Delta L$)
divided by the solid angle into which the emission is beamed,
and the density enhancement only decelerates the emitting matter
to reduce the relativistic beaming.\footnote{
If we consider a void in the ambient medium, instead of the density
enhancement, 
the matter freely expands in the void,
so that the Lorentz factor becomes higher than that in other parts.
However the difference within $\Delta t\le t$ is only a factor of $\sim 2$
and negligible for an order-of-magnitude argument.
}
We may estimate the factor $F/\nu F_{\nu}$ in equation (\ref{eq:den})
assuming the standard afterglow model as
$F/\nu F_{\nu} \sim (\nu/\nu_{c})^{(p-3)/2}$
for $\nu_{m}<\nu<\nu_{c}$ (the optical band at $t \sim 1$ day)
and  $F/\nu F_{\nu} \sim (\nu/\nu_{c})^{(p-2)/2}$
for $\nu_{m}<\nu_{c}<\nu$ (the X-ray band at $t \sim 1$ day),
since the synchrotron flux $\nu_{c} F_{\nu_{c}}$ 
at the cooling frequency $\nu_{c}$ is about the bolometric flux $F$
for $p\sim 2.2$ \citep{sariea98}.
Since $\nu_{c} \sim 10^{15}$ Hz at $t\sim 1$ day, 
we have $F/\nu F_{\nu} \sim 1$
for the optical and X-ray bands ($\nu \simg 10^{15}$ Hz).

Finally, taking the ring-like image of the afterglow into account
as in equation (\ref{eq:V-2}),
we replace equation (\ref{eq:den}) by
\beqa
\frac{\Delta F_{\nu}}{F_{\nu}}
\le \left\{
\begin{array}{ll}
(4/5) f_{c}^{-1} (F/\nu F_{\nu}) (\Delta t/t),& ({\rm on-axis})\\
6 f_{c}^{-1} (F/\nu F_{\nu}) (\Delta t/t)^{2}.& ({\rm off-axis})
\end{array}\right.
\label{eq:den2}
\eeqa
Note that the above derivation uses only the properties of the
standard afterglow. 

The above limit may be applied when the emitting site has a single 
variable region.
When many regions fluctuate simultaneously, 
we may use equation (\ref{eq:manyden}) instead, 
as discussed in \S~\ref{sec:many}.
Note, however, that the limit in equation (\ref{eq:den2}) is still useful
since its violation implies many variable regions.

\subsection{Patchy shell model}\label{sec:patchy}

An intrinsic angular structure on the emitting surface (patchy shell)
may also produce the variability of the afterglow \citep{meszaros98,kumar00a}.
Such patchy shells may be realized in the sub-jet model
\citep{yamazaki04,ioka01}. 
Since the visible size $\sim \gamma^{-1}$ grows as 
the Lorentz factor $\gamma$ drops, the observed flux varies 
depending on the angular structure.

The variability timescale is always $\Delta t \ge t$ \citep{nakar04}
if the angular fluctuation is persistent (see also \S~\ref{sec:dis}).
The rise time of the variability $\Delta t_{\rm rise}$ 
could be $\Delta t_{\rm rise}/t \sim 2 \gamma \Delta \theta < 1$
since it is determined by the timescale on which
the angular fluctuation $\Delta \theta$ enters the visible region.
(The lateral expansion has to be slow for 
$\Delta \theta < \gamma^{-1}$; see \S~\ref{sec:dis}.)
However it takes $\sim t$ for the flux to go back
to its mean level since the visible region expands on the timescale $\sim t$.

\subsection{Refreshed shocks}\label{sec:refresh}

Afterglow variabilities may also arise in the refreshed shock scenario,
in which multiple shells are ejected instantaneously
(i.e., the ejection timescale is comparable to the GRB duration and 
short compared to the observed time)
but the inner shell is so slow that it catches up with the outer shells 
on a long timescale
when the Lorentz factor of the outer shells
drops slightly below that of the slow shell
\citep{rees98,panaea98,kumar00b,sari00,zhang02}.
Since inner shells only increase the afterglow energy,
the observed flux does not go back to the original level
(i.e., the extrapolation from the original power-law).

The variability timescale should be $\Delta t \ge t/4$ 
if the acceleration of the GRB ejecta is hydrodynamic.
This is because, if hydrodynamically accelerated, the slow shell has
an opening angle larger than the inverse of its Lorentz factor 
$\Delta \theta \ge \gamma_{s}^{-1}$
and a factor of $\sim 2$ dispersion in 
the Lorentz factor of the slow shell $\gamma_{s}$,
since the shell is hot in the acceleration regime
and expands with a sound speed $\sim c$ in the comoving frame.
If $\Delta \theta \ge \gamma_{s}^{-1}$,
the variability timescale
is equal to or larger than the observed time (angular spreading time)
$\Delta t \ge R/2 c \gamma_{s}^{2} \sim t/4$,
since the Lorentz factor of the slow shell is comparable
to that of the blast wave $\gamma_{s} \sim 2 \gamma$ when they collide.
Even when $\Delta \theta < \gamma_{s}^{-1}$,
if the Lorentz factor of the slow shell $\gamma_{s}$
varies at least by a factor of $\sim 2$,
the slow shell spreads to have a width 
in the lab-frame $\Delta \sim R/2 \gamma_{s}^{2} \sim c t/4$,
so that the variability timescale is again $\Delta t \ge t/4$.
Therefore another acceleration mechanism is required for $\Delta t < t/4$.

The slow shell may satisfy $\Delta \theta < \gamma_{s}^{-1}$
if the outer shell has an opening angle smaller than $\gamma_{s}^{-1}$.
This is because a part of the slow shell
outside the wake of the outer shell is decelerated 
and cut off by the ambient material.
Only the shell in the wake remains cold and freely expands.
This mechanism may explain the bumps in GRB 030329
with $\Delta t \sim t_{j} < t$
where $t_{j}$ is the jet break time \citep{granot03}.
However the dispersion of $\gamma_{s}$
should be small for $\Delta \le c \Delta t$, 
so that a non-hydrodynamical acceleration is still needed.

\subsection{Long-acting engine model}\label{sec:long}

A bump at an observed time $t$ may
suggest that the central engine is still active at that time $t$
\citep{rees00,zhang02,dai98}.
A very long duration could arise if it takes a long time
for the disk around a black hole to be completely absorbed,
such as in some cases of the collapsar models \citep[e.g.,][]{macfadyen01}
or if the central object becomes a neutron star, such as
a millisecond magnetar \citep[e.g.,][]{usov94}.

Afterglow variabilities may arise when
a long-term intermittent outflow from the central engine
collides with the preceding blast wave,\footnote{
In the refreshed shock scenario (in the previous section),
shells are ejected at almost the same time, not for a long time.
}
with itself (internal shock),
or with the progenitor stellar envelope.
In principle, the variability timescale could be down to millisecond level
(light crossing time of the central engine)
and there is no limit on the flux variability.
The minimum energy to produce the variability is
\beqa
\frac{\Delta E}{E} \ge
\frac{\nu F_{\nu}}{F}
\frac{\Delta F_{\nu}}{F_{\nu}}
\frac{\Delta t}{t}
\frac{\Omega}{\pi \gamma^{-2}},
\eeqa
where $\Omega$ is the solid angle into 
which the variable emission is collimated and
$E$ is the afterglow energy in the visible region.
Since the solid angle $\Omega$ may be as low as 
$\sim \pi \gamma_{v}^{-2}$,
the inverse square of the Lorentz factor of the emitting matter,
there is in principle no lower limit for $\Omega$,
and hence for the minimum energy $\Delta E$,
if we consider a high Lorentz factor $\gamma_{v}$.
Considering other effects (such as multi-shock emission upon
collision) may require a larger energy \citep{zhang02}. 
High energy gamma-rays could be a diagnosis of the model
\citep{ramirez04}.

\subsection{Others}

There are several other possibilities
to produce variabilities in afterglow light curves:
the gravitational microlensing \citep{loeb98,garnavich00,ioka01b},
combined reverse shock and forward shock emission \citep{kobayashi03},
supernova bumps \citep[e.g.,][]{bloom99},
and dust echos \citep{esin00,meszaros00}.
These are not repeated,
and the variability timescales are usually $\Delta t/t \simg 0.1$.
Since these variabilities have distinct temporal and spectral features,
we will be able to distinguish these possibilities
from the mechanisms we have discussed in this paper.

\section{Solutions to forbidden variabilities}\label{sec:dis}

We have studied kinematical limits on the afterglow variabilities for
both dips and bumps, which are summarized in Figure~\ref{fig:f1}:
(a) Dips have a smaller amplitude than that given by equation (\ref{eq:V-2}),
(b) Density fluctuations cannot make a bump larger than the limit of
equation (\ref{eq:den2}),
(c) Patchy shells cannot make a bump with a timescale 
shorter than the observed time $\Delta t < t$,
(d) Refreshed shocks cannot make a bump with $\Delta t < t/4$.
If many regions fluctuate simultaneously,
the limits (a) and (b) are replaced by equations (\ref{eq:manydip}) 
and (\ref{eq:manyden}), respectively, as discussed in \S~\ref{sec:many}.
The variabilities in GRB 011211
\citep[$\Delta t/t \sim 0.1$ and $|\Delta F_{\nu}|/F_{\nu} \sim 0.1$;][]
{jakobsson04,holland02}
may violate the limits (a), (c) and (d) and marginally violate (b),
while the variabilities in GRB 021004 
\citep[$\Delta t/t \sim 0.01$ and $|\Delta F_{\nu}|/F_{\nu} \sim 0.05$;][]
{bersier03,halpern02}
may violate all these limits, 
if a single region fluctuates, although still within errors.

(1) One possible explanation to the forbidden variabilities
is the day-long central engine model discussed in \S~\ref{sec:long}.
Interestingly, this scenario may easily produce
metal features \citep{rees00}
and the metal emission lines are indeed observed
in the X-ray afterglow of GRB 011211 \citep{reeves02}.
A simple form of this model may explain forbidden bumps, but not dips.
Nevertheless, as noted by \citet{rees00},
we cannot rule out the possibility that 
all afterglow is due to the central engine itself.
This extreme version of this model may resolve 
the forbidden bumps as well as dips in the afterglow.

(2) Another solution could be provided by non-standard assumptions
that the emitting surface is anisotropic
(during the period when light curves smoothly decay)
and that the anisotropy is temporal.\footnote{
In the patchy shell model,
the anisotropy is not temporal but persistent (see \S~\ref{sec:patchy}).
}
Since equation (\ref{eq:S-}) is derived from geometrical arguments
and applicable to both dips and bumps,
the angular size of the anisotropy needs to be
\beqa
\Delta \theta \le \left\{
\begin{array}{ll}
\gamma^{-1} (\Delta t/t)^{1/2} \sim 10^{-2},& ({\rm on-axis})\\
(1/2) \gamma^{-1} (\Delta t/t) \sim 10^{-3},& ({\rm off-axis})
\end{array}\right.
\label{eq:dth}
\eeqa
and with the temporal timescale $\sim \Delta t$ 
the surface brightness in this region
should be enhanced (bumps) or reduced (dips) by
\beqa
\frac{|\Delta I_{\nu}|}{\bar I_{\nu}}=
\frac{|\Delta F_{\nu}|}{F_{\nu}}
\frac{S}{\Delta S}
\ge \left\{
\begin{array}{ll}
(|\Delta F_{\nu}|/F_{\nu}) (t/\Delta t) \sim 10,& ({\rm on-axis})\\
4 (|\Delta F_{\nu}|/F_{\nu}) (t/\Delta t)^{2} \sim 10^{3},& ({\rm off-axis})
\end{array}\right.
\label{eq:di}
\eeqa
where numerical values are for GRB 021004
($\Delta t/t \sim 0.01$ and $|\Delta F_{\nu}|/F_{\nu} \sim 0.05$)
and we have assumed $\gamma \sim 10$ at $t \sim 1$ day.
For the off-axis case,
the brightness contrast may be too large 
$|\Delta I_{\nu}|/\bar I_{\nu} \sim 10^{3}$
to be reconciled with the observed narrow distribution of 
the geometrically corrected gamma-ray energy \citep{frail01}.

Let us examine each violation one by one in this temporally anisotropic model.
To explain narrow dips violating (a), the variable region should be initially
brighter than the limit given by equation (\ref{eq:di})
and then it should be darkened, such as by the density fluctuations.
However the angular size should also satisfy equation (\ref{eq:dth}).
This size is contrary to our common belief that the initial fluctuations
with $\Delta \theta < \gamma^{-1}$ are erased during the fireball evolution
because the visible region $\sim \gamma^{-1}$ is causally connected.
The lateral expansion has to be slower than
the usual assumption (i.e., the sound speed in the local frame)
for the initial fluctuations to prevail.
[Notice that
several numerical simulations imply a slow lateral expansion
\citep{granotea01,kumar03,cannizzo04}.]
We also expect that
other variabilities due to the patchy shell effect should
appear in the light curves
when the angular fluctuations enter the visible region.

Next let us consider bumps violating the limits (b)-(d)
in the temporally anisotropic model.
To explain variabilities violating the limit (b)
in the density fluctuation scenario, 
the angular energy distribution has to be initially anisotropic
and the energetic spot needs to be brightened by the density fluctuation,
as in the dip case (a).
Again the lateral expansion has to be slow,
and we also expect other variabilities due to the patchy shell effect.
In the patchy shell model violating (c),
the anisotropy should be temporal $\sim \Delta t$,
so that an external factor such as density fluctuations may be necessary.
Also in this case the lateral expansion has to be slow.
In order for the bumps due to refreshed shocks to violate (d),
the acceleration mechanism should be non-hydrodynamical 
(see \S~\ref{sec:refresh}).
If so, the rise time $\Delta t_{\rm rise}$ may be as short as
$\Delta t_{\rm rise}/t \sim (\gamma \Delta \theta)^{2} < 1$ (on-axis),
but the flux does not return to the original level.
The step-wise light curve is a signature of the refreshed shock.

(3) The other solution could be that many regions fluctuate simultaneously.
As discussed in \S~\ref{sec:many}, the limits (a) and (b) are modified
in this case and are given by equations (\ref{eq:manydip}) 
and (\ref{eq:manyden}), respectively.
The limit (a) for dips in equation (\ref{eq:manydip})
may be still violated by GRB 011211 and GRB 021004,
while the limit (b) for bumps due to density fluctuations
in equation (\ref{eq:manyden}) may not be violated.
Even so the number of variable regions has to be larger than $\sim 10^{3}$
to reconcile the limit (b) with GRB 021004,
and this suggests that the mean separation of the density clumps
(with a radius $\sim 10^{14}$ cm) is about $\sim 10^{15}$ cm 
(see \S~\ref{sec:many}).
Therefore these limits provide interesting constraints on density fluctuations.
In this model the anisotropy of the emitting surface is also
strong and temporal as in the previous model,
and the observed flux almost always differs from the base level.

(4) The violations could be attributed to uncertainties of the
observations and data analyses.
The violation of equation (\ref{eq:V-2}), which gives
constraints on dips, may be due to
the fitting scheme, because spurious dips might be produced by 
inappropriate power-law fitting to a light curve containing bumps.
Intensive afterglow monitorings such as those by the Swift satellite
and its ground-based follow-up observations, when combined with
appropriate data fitting methods, might be able to lead to a better 
determination of the base level of the decay, which could be used to
verify or refute the presence of the forbidden variabilities. 

Some afterglows show little variabilities despite dense sampling
\citep[e.g.,][]{laursen03,gorosabel04,stanek99}.
The variety could arise from the viewing angle 
(e.g., the anisotropy depends on the viewing angle)
or the intrinsic diversity
(e.g., each burst has a different anisotropy), 
but future observations are needed
to fix its origin.

We have considered a fixed lab-frame time to relate the emitting area 
and volume with the observed time in equations (\ref{eq:S-}) and 
(\ref{eq:volume}). If we take the equal arrival time surface into account 
\citep{pana98,sari98,waxman97}, these relations would be modified and make
the limits more stringent especially near the limb of the afterglow image.
This is also an interesting future problem.

\acknowledgments
We thank P.~{\meszaros}, D.~Lazzati and the referee for useful comments.
This work was supported in part by the Eberly Research Funds of Penn State 
and by the Center for Gravitational Wave Physics under grants 
PHY-01-14375 (for KI, SK),
NASA Swift Cycle 1 GI Program (for SK, BZ) and
NASA NNG04GD51G (for BZ).

\appendix

\section{The case of many variable regions}\label{sec:many}

So far we have considered only a single variable region to make the 
discussions simple. In reality many regions could fluctuate 
simultaneously \citep[e.g.,][]{gruzinov99}. This leads to weaker 
limits for dips (a) and bumps due to density fluctuations (b) 
in equations (\ref{eq:V-2}) and (\ref{eq:den2}), respectively. 
In any case, we can still obtain meaningful limits by 
extending the previous arguments, as shown below.

Let $N_{v}$ be the mean number of variable regions contributing
in the observed time interval of $\sim \Delta t$.
Then the flux deviates from the baseline
$\sim N_{v}$ times larger than the single variable region case.
However the flux variability is 
determined by the differences between time intervals.
This is determined by the variance of the number of variable regions
from one time interval $\sim \Delta t$ to the next,
and it is $\sim N_{v}^{1/2}$ if we assume the Poisson statistics.
The Poisson statistics may be applied
since different variable regions fluctuate at different lab-frame times.
Therefore the limits for a single variable region
in equations (\ref{eq:V-2}) and (\ref{eq:den2})
have to be multiplied by $\sim N_{v}^{1/2}$ 
when many regions fluctuate.

\subsection{Limits (a): dips}

Let us consider the maximum of the number of variable regions $N_{v}$ 
in the dip case (a).
In the on-axis case, the maximum number is about the ratio of the overall
thickness of the blast wave $\sim R/16\gamma^{2} \sim ct/8$ 
to the fluctuating thickness $\sim c\Delta t$,
i.e., $N_{v} \le (t/\Delta t)/8$.
In the off-axis case, it is about the ratio of the visible volume $V$
to the variable volume $\Delta V$ in equation (\ref{eq:volume}), i.e., 
$N_{v} \le V/\Delta V \sim (t/\Delta t)^{3}/2$.

Then, multiplying the limits for a single variable region
in equation (\ref{eq:V-2}) by $\sim N_{v}^{1/2}$,
we obtain the limits for dips due to many variable regions as
\beqa
\frac{|\Delta F_{\nu}|}{F_{\nu}}
\le \left\{
\begin{array}{ll}
(\sqrt{2}/5)(\Delta t/t)^{3/2},& ({\rm on-axis})\\
(6/\sqrt{2}) (\Delta t/t)^{3/2}.& ({\rm off-axis})
\end{array}\right.
\label{eq:manydip}
\eeqa
The off-axis limit is shown by the dashed line in Figure~\ref{fig:f1}.
The variabilities in GRB 011211
($\Delta t/t \sim 0.1$ and $|\Delta F_{\nu}|/F_{\nu} \sim 0.1$)
and GRB 021004 
($\Delta t/t \sim 0.01$ and $|\Delta F_{\nu}|/F_{\nu} \sim 0.05$)
may still violate these limits.

The above limits are quite robust, but may be too strict
because the interior of the blast wave also needs to fluctuate 
in the same pace as the front and back of the blast wave
when the number of variable regions $N_{v}$ is nearly maximum.
A reasonable fluctuating site may be the front and back of the blast wave
that are affected by the density clumps or the inner shells.
If this is the case, the maximum of $N_{v}$ is $\sim 1$ in the on-axis case,
and is about the ratio of the visible area $S$
to the variable area $\Delta S$ in equation (\ref{eq:S-}), i.e., 
$N_{v} \le S/\Delta S \sim 4 (t/\Delta t)^{2}$, in the off-axis case.

\subsection{Limits (b): bumps due to ambient density fluctuations}

When the density fluctuations make bumps in the afterglow,
the bumps are mainly produced at the shock front
(i.e., not such as in the interior of the blast wave).
Then the maximum of the number of variable regions $N_{v}$ 
is $\sim 1$ in the on-axis case.
In the off-axis case, it is about the ratio of the visible area $S$
to the variable area $\Delta S$ in equation (\ref{eq:S-}), i.e., 
$N_{v} \le S/\Delta S \sim 4 (t/\Delta t)^{2}$.

\begin{figure}
\plotone{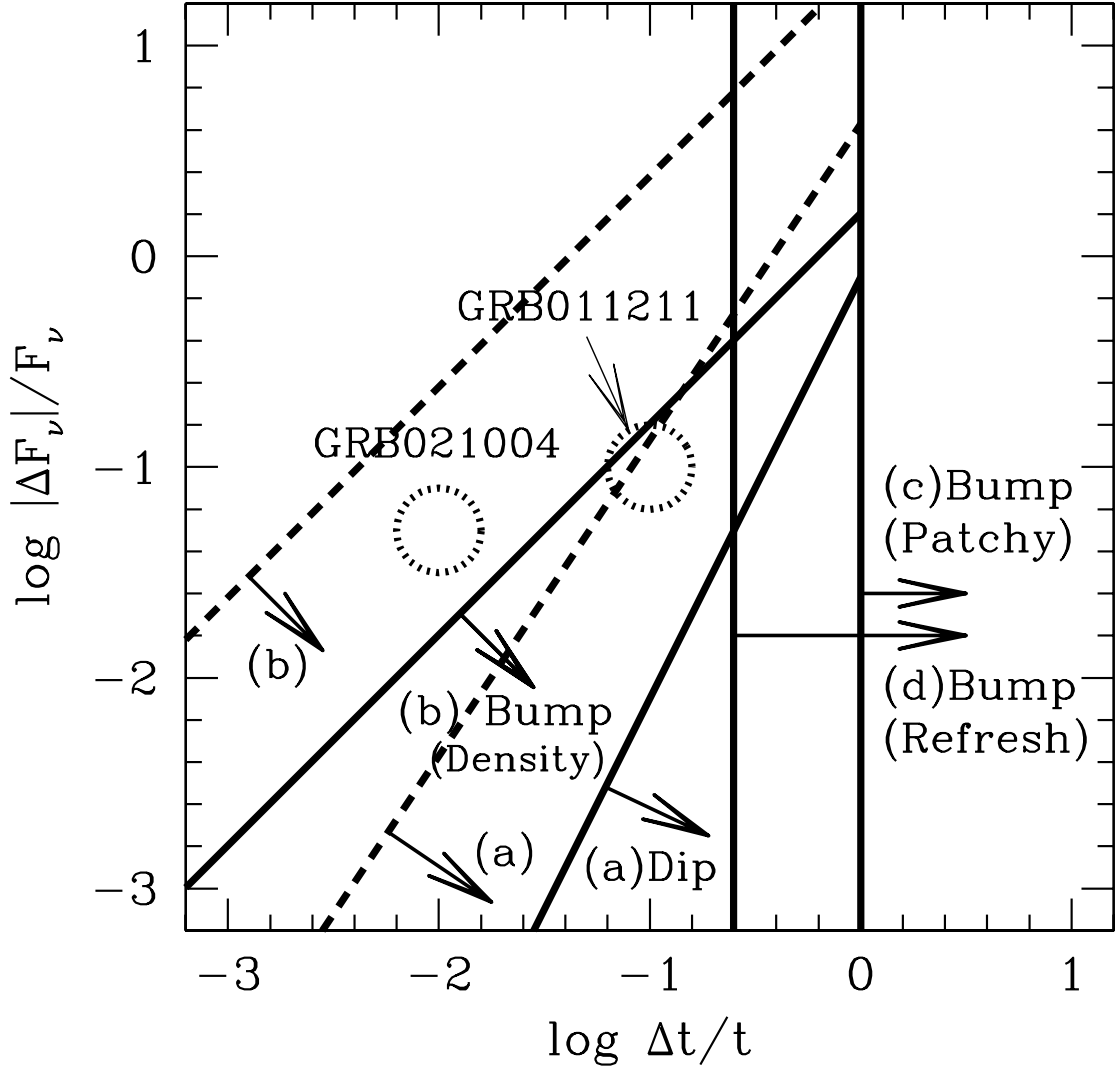}
\caption{\label{fig:f1}
Kinematically allowed regions for afterglow variabilities
are shown in the plane of the relative variability timescale
$\Delta t/t$ and relative variability flux $\Delta F_{\nu}/F_{\nu}$.
We have four limits:
(a) equation (\ref{eq:V-2}) for dips, 
(b) equation (\ref{eq:den2}) for bumps due to density fluctuations,
(c) $\Delta t \ge t$ for bumps due to patchy shells
and (d) $\Delta t \ge t/4$ for bumps due to refreshed shocks.
For (a) and (b), the on-axis cases are shown.
When many regions fluctuate simultaneously,
the limits (a) and (b) are replaced by 
equations (\ref{eq:manydip}) and (\ref{eq:manyden}),
respectively, and the off-axis cases are shown by dashed lines.
We assume $F/\nu F_{\nu} \sim 1$ and $f_{c}\sim 1/2$.
The variabilities in the afterglows of GRB 011211 \citep{jakobsson04,holland02}
and GRB 021004 \citep{bersier03,halpern02}
are also shown by dotted circles, and may violate some of these limits.
}
\end{figure}

Multiplying the limits for a single variable region
in equation (\ref{eq:den2}) by $\sim N_{v}^{1/2}$,
we obtain the limits for bumps due to many density fluctuations as
\beqa
\frac{\Delta F_{\nu}}{F_{\nu}}
\le \left\{
\begin{array}{ll}
(4/5) f_{c}^{-1} (F/\nu F_{\nu}) (\Delta t/t),& ({\rm on-axis})\\
12 f_{c}^{-1} (F/\nu F_{\nu}) (\Delta t/t).& ({\rm off-axis})
\end{array}\right.
\label{eq:manyden}
\eeqa
The off-axis limit is shown by the dashed line in Figure~\ref{fig:f1}.
This maximum off-axis limit is not violated by
the variabilities in GRB 011211
($\Delta t/t \sim 0.1$ and $|\Delta F_{\nu}|/F_{\nu} \sim 0.1$)
and GRB 021004 
($\Delta t/t \sim 0.01$ and $|\Delta F_{\nu}|/F_{\nu} \sim 0.05$).
Still we need $N_{v} \simg 1600$ variable regions
to reconcile the off-axis limit with GRB 021004
because the variabilities in GRB 021004 are
$\sim 40 \sim \sqrt{1600}$ times larger than the off-axis limit 
for a single variable region in equation (\ref{eq:den2}).
This corresponds to the mean separation of the density clumps
$\sim [\pi (R/\gamma)^{2} c \Delta t \gamma^{2}/N_{v}]^{1/3} \sim 10^{15}$ cm
since the shock front with a radius $R/\gamma \sim c t \gamma \sim 10^{16}$ cm
sweeps a distance $c \Delta t \gamma^{2} \sim 10^{15}$ cm
for $t \sim 1$ day, $\Delta t \sim 10^{3}$ sec and $\gamma \sim 10$.
Note that we can also estimate the clump radius as
$c \Delta t \gamma \sim 10^{14}$ cm.
In this way, the limits in equations (\ref{eq:den2}) and (\ref{eq:manyden})
can impose interesting constraints on the density fluctuations.

%
%

%
%

\newpage

\end{document}